\documentclass[prb,preprint,a4paper,showpacs,superscriptaddress]{revtex4-1}
\usepackage{graphicx}
\usepackage{bm}
\usepackage{color}

\renewcommand{\k}{\mathbf{k}}

\begin{document}

\title{Noise Spectra of ac-driven quantum dots}

\author{B. H. Wu}

\email{bhwu@mail.sim.ac.cn}
\affiliation{Max Planck Institute for
the Physics of Complex Systems, N\"{o}thnitzer Str.\ 38, 01087
Dresden, Germany} \affiliation{State Key Laboratory of Functional
Materials for Informatics, Shanghai Institute of Microsystem and
Information Technology, 865 Changning Road, Shanghai 200050, China}
\author{C. Timm}
\email{carsten.timm@tu-dresden.de}
\affiliation{Institute for Theoretical Physics, Technische
Universit\"at Dresden, 01062 Dresden, Germany}

\begin{abstract}
We study the transport properties of a quantum dot driven by either
a rotating magnetic field or an ac gate voltage using the Floquet
master-equation approach. Both types of ac driving lead to
photon-assisted tunneling where quantized amounts of energy are
exchanged with the driving field. It is found that the
differential-conductance peak due to photon-assisted tunneling
does not survive in the Coulomb-blockade regime when the dot is driven
by a rotating magnetic field. Furthermore, we employ a generalized
MacDonald formula  to calculate the time-averaged noise spectra of
ac-driven quantum dots. Besides the peak at zero frequency, the
noise spectra show additional peaks or dips in the presence
of an ac field. For the case of an applied ac gate voltage,
the peak or dip position is fixed at the driving frequency,
whereas the position changes with increasing amplitude for the case of a
rotating magnetic field. Additional features appear in the noise spectra if a
dc magnetic field is applied in addition to a rotating field. In all cases,
the peak or dip positions can be understood from the energy differences
of two available Floquet channels.
\end{abstract}


\maketitle

\section{Introduction}

Quantum conductors based on single molecules or semiconductor
quantum dots are promising building blocks for future electronics
and model systems for the study of fundamental quantum
phenomena.\cite{PT5643}
However, much information on quantum
conductors is beyond the reach of measurements of the
current or conductance alone. Instead, the understanding of the
transport properties calls for a study of the full counting
statistics.\cite{JMP374845,PRB70205334,PRL99076601,Nazarov,PRL94206804,PRB74205438,JPCM21295602,Nature434361,Science3121634,PRL96076605}
In the past decade, valuable information on microscopic details
of the charge transport has been obtained from measurements of the
current fluctuations or current noise.\cite{PR3361} Previous studies
have shown that one can extract parameters such as the average
backscattered charge,\cite{PRB77241303} the intrinsic time
scales,\cite{PRL92206601,PRL96056603} and the asymmetry of the dot-lead
coupling\cite{PRB68115105,PRB79075307} from current-noise measurements.
Most studies of the current noise have focused
on the zero-frequency noise
power $S(0)$.\cite{PRL90210602,PRB77233307,arxiv09065266}   The
zero-frequency noise reflects the average properties of the
tunneling. Since the finite-frequency current noise $S(\omega)$ is a
measure of the correlations between tunneling events with their time
difference conjugate to the frequency $\omega$,\cite{PRB74075328} it is
interesting to go beyond the zero-frequency limit. Aguado and
Brandes \cite{PRL92206601} have demonstrated that the noise spectra
can show dip structures at the splitting energy of an open quantum
two-level system, with their width controlled by its dissipative
dynamics.
In many works, the MacDonald formula\cite{RPP1256} has
been used to study the current-noise
spectra.\cite{PRB76085325,PRB75045340,JAP104033532}

Effective \textit{in-situ} manipulation of quantum conductors is a key
step for further development. Using a time-dependent field
to manipulate the dynamics of quantum dots promises to be
advantageous in situations ranging from photon-assisted inelastic
tunneling\cite{PR129647} to quantum pumping.\cite{PRB58R10135}
When the conductor is driven by an ac field, one expects
novel features due to the interplay of intrinsic oscillation frequencies and
the external driving frequency.
Several recent studies\cite{PRL96017405,PRL102016802}
indicate that key information is hidden in the noise spectra
of the ac-driven transport. For instance, Barrett and
Stace\cite{PRL96017405} have proposed to extract the characteristic
timescales such as the inverse dephasing and relaxation rates of a solid-state
charge qubit coupled to a microwave field from the noise spectrum. Wabnig
\emph{et al.}\cite{PRL102016802} have proposed
to estimate the coherence time of the spin in a quantum dot by
measuring its noise spectra under an ac magnetic field.  These
results are obtained based on the rotating-wave approximation and
usually in the limit of infinite on-site Coulomb interaction.

For periodically driven systems, an appropriate theoretical tool to go
beyond the rotating-wave approximation
is the Floquet theorem.\cite{Floquet,PR304229}
Various attempts have been
made by generalizing the existing steady-state transport approaches
such as the scattering
matrix\cite{PRB66205320,PRB73245312,PRB77035329} and nonequilibrium
Green's functions\cite{PRB72245339,PRB72125349,JPCM20085224}
with the help of the Floquet theorem.
However, these methods are not adequate to fully take
the Coulomb blockade in quantum dots into account, which
dominates the transport properties of small-size quantum conductors.
The quantum master equation\cite{Breuer} in its various
manifestations\cite{PRB77195416,PRB71155403} is able to give a good
account of the Coulomb blockade in the weak-tunneling limit. This
method has previously been generalized
using the Floquet theorem to study the current and the zero-frequency
noise power in an ac-driven
conductor.\cite{PRA444521,PRE614883,Breuer,JCP1212278,AP16702,arxiv09041687}

In the present study, we employ the Floquet master equation in the Fock
space of an ac-driven quantum dot to study the transport properties
such as the differential conductance and the full noise spectrum in the
sequential-tunneling limit. As the ac field we consider a rotating magnetic
field as well as an ac gate voltage for comparison. We employ a
generalized MacDonald formula for the time-averaged noise spectra in the
presence of a periodic ac field. An equivalent form of the
generalized MacDonald formula has been given by Clerk
and Girvin\cite{PRB70121303} without derivation. For completeness, we
present a derivation in the appendix. Note that the authors
are concerned with a different case, namely an ac \emph{bias} voltage, and do
not employ the Floquet formalism. 

Our paper is organized as follows. In Sec.\ \ref{sec:formal},
a Floquet master-equation
formalism is presented to study the transport properties of
ac-driven quantum dots. Expressions for the noise spectra are derived
based on the
full counting statistics and the generalized MacDonald formula. In
Sec.\ \ref{sec:results},
the transport properties of the ac-driven quantum dot are studied.
The ac field is either a rotating magnetic
field or an oscillating on-site energy due to a periodic gate voltage.
The characteristic features in the transport properties are
presented and discussed. In Sec.\ \ref{sec:sum}, a brief summary
is given.

\section{Formalism}
\label{sec:formal}

\subsection{Model}

In this paper, we study the transport properties of a single-level
quantum dot driven by an ac field. The quantum dot is coupled to
the left ($L$) and right ($R$) electron leads. The leads are assumed
to be ideal and free of interactions. The Hamiltonian of the model
system can be written as
\begin{eqnarray}
  H(t)=H_{L}+H_R+H_{\mathrm{dot}}(t)+H_T
  \equiv H_0 + H_T ,
\end{eqnarray}
where $H_{\mathrm{dot}}$ is the Hamiltonian of the isolated quantum
dot, which contains the effects of the ac field and the Coulomb
interaction,
$H_l=\sum_{\k\sigma}\epsilon_{l\k\sigma}c^\dag_{l\k\sigma}c_{l\k\sigma}$
represents the Hamiltonian of lead $l=L,R$,
where $c_{l\k\sigma}$ ($c^\dag_{l\k\sigma}$) annihilates (creates)
an electron with spin $\sigma$, crystal momentum $\k$,
and energy $\epsilon_{l\k\sigma}$ in lead $l$, and
$H_T=\sum_{l\k\sigma}V_{l\k\sigma}c^\dag_{l\k\sigma}d_\sigma+\mathrm{h.c.}$
describes the coupling between the quantum dot and the
leads, where $d^\dag_\sigma$ ($d_\sigma$) is the spin-$\sigma$
electron creation (annihilation) operator in the quantum dot.
We note that we do not assume an infinite Coulomb-interaction
strength $U\to\infty$, in contrast to previous studies.\cite{AP16702}
Instead, finite Coulomb
interaction will be included by taking the doubly
occupied state into account.

In the following, we focus on the limit of weak dot-lead coupling
and investigate the transport properties of the quantum dot using
the Floquet master-equation method. For a small quantum dot, for
which the Coulomb interaction can dominate the transport behavior,
the treatment of the Coulomb interaction must go beyond the
mean-field level.  To this end, it is convenient to rewrite the dot
Hamiltonian in the electron-number basis of the Fock
space.\cite{Datta_arxiv,PRB79205303} In this description, the
quantum dot can either be in the empty state $|0\rangle$, the singly
occupied state $|\sigma\rangle$ with spin $\sigma=\uparrow$ or
$\downarrow$, or the doubly occupied state
$|\uparrow\downarrow\rangle$. In the following, we denote the states
in the Fock space by Latin letters, $|a\rangle=|0\rangle,
|\uparrow\rangle, |\downarrow\rangle, |\uparrow\downarrow\rangle$.
Using this orthonormal basis, the dot-lead coupling can be described
naturally  with the help of Hubbard operators
\begin{eqnarray}
  X_{ab}=|a\rangle\langle b|,
\end{eqnarray}
which describes the transition of the quantum dot from state
$|b\rangle$ to state $|a\rangle$. The second-quantized dot-electron
creation operator can thus be rewritten in terms of the Hubbard
operators as
\begin{eqnarray}
  d^\dag_\sigma=|\sigma\rangle\langle
  0|+\eta_\sigma|\uparrow\downarrow\rangle\langle\bar\sigma|,
\end{eqnarray}
where $\bar\sigma$ represents the opposite spin of $\sigma$ and the
factor $\eta_\sigma=\pm 1$ for $\sigma=\uparrow,\downarrow$,
respectively, is due to the anticommutation relation of the
Fermions. In terms of these Hubbard operators, the Hamiltonian for
the dot-lead coupling and the isolated dot can be rewritten as
\begin{eqnarray}
  H_\mathrm{dot}&=&\sum_{ab}H^D_{ab}|a\rangle\langle b|,\\
 H_T&=&\sum_{ab,l\k\sigma}V^{ab}_{l\k\sigma} c^\dag_{l\k\sigma}
X_{ab}+\mathrm{h.c.},
\end{eqnarray}
respectively. Here, we have made the coupling strength
$V^{ab}_{l\k\sigma}$ depend on the occupancy of the initial and
final states of the transition in the quantum dot. The explicit form
of the dot Hamiltonian depends on the details of the device geometry
and the external ac driving field. It will be specified in the
following sections.

\subsection{The Floquet quantum-master-equation approach}

\subsubsection{Floquet states}

Due to the presence of a time-periodic external field, the dynamics
of the quantum dot is governed by a Hamiltonian that is periodic in
time with the frequency $\Omega=2\pi/\mathcal{T}$, i.e.,
$H_\mathrm{dot}(t)=H_\mathrm{dot}(t+\mathcal{T})$, where
$\mathcal{T}$ denotes the period. The solution of the time-periodic
Hamiltonian can be simplified by the Floquet
theorem,\cite{Floquet,PR304229} which states that the solution of
the Schr\"odinger equation for the dot Hamiltonian can be obtained
from (we set $\hbar=|e|=k_B=1$ in the following)
\begin{eqnarray}\label{Floq.eq1}
  \left[H_\mathrm{dot}(t)-i\,\frac{\partial}{\partial
  t}\right]|\alpha (t)\rangle=\varepsilon_\alpha\,|\alpha(t)\rangle,
\end{eqnarray}
where $\varepsilon_\alpha$ is the time-independent Floquet quasienergy and
$|\alpha(t)\rangle$ is the corresponding Floquet state, which has
the same period $\mathcal{T}$,
$|\alpha(t)\rangle=|\alpha(t+\mathcal{T})\rangle$. Here, Greek
letters are used to denote the Floquet states. Further simplification is
possible by decomposing the Floquet states into a Fourier series,
\begin{eqnarray}
  |\alpha(t)\rangle&=&\sum_k e^{-ik\Omega t}\, |\alpha_k\rangle ,
\end{eqnarray}
with the reverse transformation
\begin{eqnarray}
  |\alpha_k\rangle&=&\frac{1}{\mathcal{T}}\int^\mathcal{T}_0dt\,
  e^{ik\Omega t}\, |\alpha(t)\rangle
\end{eqnarray}
and analogously for $H_\mathrm{dot}(t)$.
The Fourier transform of Eq.\ (\ref{Floq.eq1}) then reads
\begin{eqnarray}
\sum_{k'} H_{\mathrm{dot},k-k'}\, |\alpha_{k'}\rangle
  - k \Omega \, |\alpha_k\rangle
  &=& \varepsilon_\alpha\, |\alpha_k\rangle .
\end{eqnarray}
The quasienergy $\varepsilon_\alpha$ can evidently be restricted to the first
Brillouin zone $[0,\Omega)$ of the Floquet space, while the Floquet index $k$
can assume any integer value. Equivalently, we can view
$\varepsilon_\alpha+k\Omega$ as the quasienergy in the extended zone
scheme.

We also introduce the Hubbard operator in the Floquet
states to describe the transition between the Floquet states as
$X_{\alpha\beta}(t)=|\alpha(t)\rangle\langle\beta(t)|$. For the
time-dependent transport, it is more convenient to work with these
Floquet states. This is most advantageous in transformations of the following
form, which we will use in the derivation below,
\begin{eqnarray}\label{UXU}
\tilde
X_{\alpha\beta}(t',t)&=&U^\dag_0(t',t)\,
  X_{\alpha\beta}(t')\,U_0(t',t) \nonumber \\ &=&
  e^{i(\varepsilon_\beta-\varepsilon_\alpha)(t-t')}X_{\alpha\beta}(t),
\end{eqnarray}
where
\begin{equation}
U_0(t',t) =
T_c\exp\left(-i\int^{t'}_{t}dt''\,  [H_L+H_R+H_\mathrm{dot}
(t'')]\right)
\end{equation}
denotes the time-evolution operator due to the Hamiltonian
in the absence of tunneling. Here, $T_c$ is the time-ordering
operator and the dot Hamiltonian is explicitly time-dependent.

\subsubsection{The Floquet quantum master equation with counting fields}

For a quantum dot coupled to external leads, the exact quantum master equation
can be written in the interaction picture as\cite{Breuer,PRB77195416}
\begin{eqnarray}\label{EOMrhoI}
  \frac{d}{dt}\rho_I(t)=-i[H_{T,I}(t),
  \rho_I(t_0)]-\int_{t_0}^tdt'\,[H_{T,I}(t), [H_{T,I}(t'),\rho_I(t')]],
\end{eqnarray}
where $A_I(t)=U^\dag_0(t,t_0)A(t)U_0(t,t_0)$ denotes an operator in the
interaction picture and $\rho(t)$ is the density matrix in the Fock
space of the full system.

A complete description of the electronic transport through the quantum
dot is provided by the full counting statistics.
Properties such as the noise spectrum are determined by the
counting statistics of the electrons arriving at and departing from
the leads. All information on the counting statistics is contained in
the moment-generating function $\phi(\chi_L,\chi_R)=\langle
\exp(i\chi_LN_L+i\chi_RN_R)\rangle$. Here, $\chi_l$ represents the
counting field in the lead $l$, which counts how many electrons
have tunneled into or out of the lead.
$N_l=\sum_{\k\sigma}c^\dag_{l\k\sigma}c_{l\k\sigma}$ is the
electron-number operator in lead $l$. We introduce the operator
\begin{eqnarray}
  \mathcal{F}(\chi_L,\chi_R,t)=
  \mathrm{Tr}_{\mathrm{leads}}\, e^{i\chi_LN_L+i\chi_RN_R}\rho(t) .
\end{eqnarray}
In this we follow Kaiser and
Kohler,\cite{AP16702} except that we introduce two counting fields.
In the limit of $\chi_L\rightarrow 0$ and $\chi_R\rightarrow 0$, $\mathcal{F}$
becomes the reduced density matrix of the quantum dot,
$\rho_\mathrm{dot}=\mathrm{Tr}_\mathrm{leads}\,\rho$. Moreover, the
moment-generating function $\phi(\chi_L,\chi_R, t)$ can be obtained
by tracing out the dot degrees of freedom,
$\phi=\mathrm{Tr}_\mathrm{dot}\,\mathcal{F}$. We decompose
$\mathcal{F}$ into a Taylor series,
\begin{eqnarray}\label{TSF}
  \mathcal{F}=\sum_{m=0}^\infty\sum_{n=0}^{\infty}
  \frac{(i\chi_L)^{m}(i\chi_R)^{n}}{m!n!}\,\mathcal{F}^{m,n},
\end{eqnarray}
where the coefficients
\begin{eqnarray}
\mathcal{F}^{m,n}
  =\left.\frac{\partial^{m+n}}{\partial^m(i\chi_L)\partial^n(i\chi_R)}\,
  \mathcal{F}\,\right|_{\chi_L,\chi_R\rightarrow
  0}=\mathrm{Tr}_{\mathrm{leads}}\, N_L^{m}N_R^n\,\rho
\label{Fmn}
\end{eqnarray}
provide a direct
access to the moments $\langle
N_L^mN_R^n\rangle=\mathrm{Tr_{dot}}\,\mathcal{F}^{m,n}$. In
particular, we obtain the reduced density matrix of the quantum dot,
$\rho_{\mathrm{dot}}=\mathcal{F}^{0,0}$.

To find the solutions
for $\mathcal{F}$, we first transform the equation of motion for the
density matrix in the interaction picture, Eq.\ (\ref{EOMrhoI}), back to
the Schr\"odinger picture,
\begin{eqnarray}\label{QME1}
\lefteqn{ \frac{d\rho(t)}{dt}+i[H_0(t),\rho(t)]=-i[H_T,
  U^\dag_0(t_0,t)\rho(t_0)U_0(t_0,t)] } \nonumber \\
&& {}-\int_{t_0}^t dt'\, [H_T,
  U_0^\dag(t',t) [H_T,\rho(t')] U_0(t',t)] .
\end{eqnarray}
Then, we multiply by
$e^{i\chi_LN_L+i\chi_RN_R}$ from
the left and take the trace over the lead degrees of freedom to obtain
\begin{eqnarray}\label{QME2pre}
\lefteqn{
  \frac{d\mathcal{F}(\chi_L,\chi_R, t)}{dt}
  +i[H_\mathrm{dot}(t), \mathcal{F}(\chi_L, \chi_R,
  t)] } \nonumber \\
&&  = -i\, \mathrm{Tr_{leads}}\, e^{i\chi_LN_L+i\chi_RN_R}\,
  [H_T,  U^\dag_0(t_0,t)\rho(t_0)U_0(t_0,t)] \nonumber \\
&& \quad {}-\int_{t_0}^t dt'\, \mathrm{Tr_{leads}}\, e^{i\chi_L
  N_L+i\chi_RN_R} [H_T, U_0^\dag(t',t) [H_T,\rho(t')] U_0(t',t) ] ,
\end{eqnarray}
which is still exact.

We now assume that  the full density operator is of product form at
the initial time $t_0$,
$\rho(t_0)=\rho_\mathrm{dot}(t_0) \otimes \rho^0_\mathrm{leads}$, where
$\rho_{\mathrm{leads}}^0$ describes the leads in separate thermal equilibrium.
This assumption is reasonable since we are not interested in transient
effects coming from the initial state. Such effects have been
studied by Flindt \textit{et al.}\cite{PRL100150601} 
The first term on the right-hand side of Eq.\ (\ref{QME2pre}) then vanishes.
Furthermore, we make the sequential-tunneling approximation appropriate for
weak tunneling, i.e., we treat the tunneling perturbatively to second order in
$H_T$. Since two powers of $H_T$ are already explicit in the second term on
the right-hand side of Eq.\ (\ref{QME2pre}), we can express $\rho(t')$ in terms
of the unperturbed time evolution, $\rho(t') \approx U_0^\dag(t,t') \rho(t)
U_0(t,t')$.  This makes the master equation local in time, i.e.,
Markovian. For details, see, e.g., Ref.\ \onlinecite{PRB77195416}.
We thus do not include non-Markovian effects as studied by Flindt
\textit{et al.}\cite{PRL100150601} This is valid if the relaxation time in the
leads is short compared to the typical timescales of the dot, which in our
case include the period of the ac field. Since relaxation times in metallic
leads are of the order of femtoseconds, this is easily satisfied.

Finally, we send $t_0\to-\infty$ and obtain
\begin{eqnarray}\label{QME2}
\lefteqn{
  \frac{d\mathcal{F}(\chi_L,\chi_R, t)}{dt}
  +i[H_\mathrm{dot}(t), \mathcal{F}(\chi_L, \chi_R,
  t)] } \\ \nonumber
&&  =-\int_0^\infty d\tau\, \mathrm{Tr_{leads}}\, e^{i\chi_L
  N_L+i\chi_RN_R}[H_T, [\tilde H_T(t-\tau, t),\rho(t)]] ,
\end{eqnarray}
where $\tilde H_T(t',t)=U^\dag_0(t',t)H_T U_0(t',t)$.

To obtain the Floquet master equation, we write Eq.\ (\ref{QME2}) in
the basis of Floquet states $|\alpha(t)\rangle$, $|\beta(t)\rangle$.
By making use of the relation Eq.\ (\ref{UXU}) and tracing
out the lead degrees of freedom, we arrive at the equation of
motion for $\mathcal{F}$,
\begin{eqnarray}\label{EOMF}
\lefteqn{\frac{d}{dt}\mathcal{F}_{\alpha\beta}(\chi_L,\chi_R,t)=\Big\{\Big[
\mathcal{L}
  +(e^{i\chi_L}-1)\mathcal{J}_{L+}+(e^{-i\chi_L}-1)\mathcal{J}_{L-}} \nonumber
  \\
& & {}+(e^{i\chi_R}-1)\mathcal{J}_{R+}+(e^{-i\chi_R}-1)\mathcal{J}_{R-}\Big]
  \, \mathcal{F}(\chi_L,\chi_R,t)\Big\}_{\alpha\beta},\hspace{2em}
\end{eqnarray}
where the superoperators are given by
\begin{eqnarray}
\lefteqn{ (\mathcal{LF})_{\alpha\beta}
= -i(\varepsilon_\alpha-\varepsilon_\beta)\mathcal{F}_{\alpha\beta}
+\frac{1}{2\pi}\int^\infty_0d\tau\int
d\varepsilon\,\Bigg\{\sum_{ab;mn}\sum_{\gamma\delta}\sum_{l={L,R};\sigma}} \\
\nonumber
&&\left[-(f_l(\varepsilon)e^{i\varepsilon\tau}\Gamma^{l\sigma}_{ab;mn}(\varepsilon)+\bar{f}_l(\varepsilon)e^{-i\varepsilon\tau}
\Gamma^{l\sigma}_{mn;ab})\,
e^{i(\varepsilon_\delta-\varepsilon_\gamma)\tau}
\langle\alpha(t)|a\rangle\langle
b|\gamma(t)\rangle\langle\gamma(t-\tau)|m\rangle \langle
n|\delta(t-\tau)\rangle\mathcal{F}_{\delta\beta}(\chi_L,\chi_R,t)\right.\\
\nonumber
&&\left.+(\bar{f}_l(\varepsilon)e^{i\varepsilon\tau}\Gamma^{l\sigma}_{ab;mn}
(\varepsilon)
+f_l(\varepsilon)e^{-i\varepsilon\tau}\Gamma^{l\sigma}_{mn;ab})\,
e^{i(\varepsilon_\beta-\varepsilon_\delta)\tau}
\langle\alpha(t)|a\rangle\langle b|\gamma(t)\rangle
\mathcal{F}_{\gamma\delta}(\chi_L,\chi_R,t)\langle
\delta(t-\tau)|m\rangle\langle n|\beta(t-\tau)\rangle\right.\\
\nonumber
&&\left.+(\bar{f}_l(\varepsilon)e^{-i\varepsilon\tau}\Gamma^{l\sigma}_{ab;mn}
(\varepsilon)
+f_l(\varepsilon)e^{i\varepsilon\tau}\Gamma^{l\sigma}_{mn;ab})\,
e^{i(\varepsilon_\gamma-\varepsilon_\alpha)\tau}
\langle\alpha(t-\tau)|a\rangle\langle
b|\gamma(t-\tau)\rangle\mathcal{F}_{\gamma\delta}(\chi_L,\chi_R,
t)\langle\delta(t)|m\rangle \langle n|\beta(t)\rangle\right.\\
\nonumber
&&\left.-(\bar{f}_l(\varepsilon)e^{i\varepsilon\tau}\Gamma^{l\sigma}_{mn;ab}
(\varepsilon)+f_l(\varepsilon)e^{-i\varepsilon\tau}
\Gamma^{l\sigma}_{ab;mn})\,
e^{i(\varepsilon_\delta-\varepsilon_\gamma)\tau}\mathcal{F}_{\alpha\gamma}(\chi_L,\chi_R,t)
\langle\gamma(t-\tau)|a\rangle\langle
b|\delta(t-\tau)\rangle\langle\delta(t)|m\rangle\langle
n|\beta(t)\rangle\right]\Bigg\},
\end{eqnarray}
\begin{eqnarray}
\lefteqn{(\mathcal{J}_{l+}\mathcal{F})_{\alpha\beta} =
\frac{1}{2\pi}\int^\infty_0d\tau\int d\varepsilon
\sum_{ab;mn}\sum_{\gamma\delta}\sum_{\sigma}
 \bar{f}_l(\varepsilon)\Gamma^{l\sigma}_{ab;mn} } \nonumber \\
 &&\left( e^{i\varepsilon\tau}e^{i(\varepsilon_\beta-\varepsilon_\delta)\tau}
 \langle\alpha(t)|a\rangle\langle b|\gamma(t)\rangle\mathcal{F}_{\gamma\delta}(\chi_L,\chi_R,t)\langle\delta(t-\tau)|m\rangle
 \langle n|\beta(t-\tau)\rangle\right. \nonumber  \\
 &&\left.+e^{-i\varepsilon\tau}e^{i(\varepsilon_\gamma-\varepsilon_\alpha)\tau}\langle\alpha(t-\tau)|a\rangle\langle
 b|\gamma(t-\tau)\rangle\mathcal{F}_{\gamma\delta}(\chi_L,\chi_R,t)\langle\delta(t)|m\rangle\langle
 n|\beta(t)\rangle\right),
\end{eqnarray}
\begin{eqnarray}
\lefteqn{(\mathcal{J}_{l-}\mathcal{F})_{\alpha\beta} =
\frac{1}{2\pi}\int^\infty_0d\tau\int
d\varepsilon\sum_{ab;mn}\sum_{\gamma\delta}\sum_{\sigma}
f_l(\varepsilon)\Gamma^{l\sigma}_{mn;ab} } \nonumber \\
&&\left(e^{-i\varepsilon\tau}e^{i(\varepsilon_\beta-\varepsilon_\delta)\tau}
\langle\alpha(t)|a\rangle\langle
b|\gamma(t)\rangle\mathcal{F}_{\gamma\delta}(\chi_L,\chi_R,t)\langle\delta(t-\tau)|
m\rangle\langle n|\beta(t-\tau)\rangle\right.  \nonumber \\
&&\left.+e^{i\varepsilon\tau}e^{i(\varepsilon_\gamma-\varepsilon_\alpha)\tau}\langle\alpha(t-\tau)|a\rangle\langle
b|
\gamma(t-\tau)\rangle\mathcal{F}_{\gamma\delta}(\chi_L,\chi_R,t)\langle\delta(t)|m\rangle\langle
n|\beta(t)\rangle\right).
\end{eqnarray}
Here, we have defined the tunneling rate
$\Gamma^{l\sigma}_{mn;ab}(\epsilon)={2\pi}\rho_l(\epsilon)V^{ab}_{l,k,\sigma}V^{mn*}_{l,k,\sigma}$,
where $\rho_l(\epsilon)$ is the density of states in lead $l$.

Inserting the Taylor expansion of $\mathcal{F}$ in Eq.\
(\ref{TSF}) into its equation of motion [Eq.\ ($\ref{EOMF}$)], one obtains
a hierarchy of equations for the expansion coefficients,
\begin{eqnarray}
  \frac{d}{dt}\mathcal{F}^{0,0}&=&\mathcal{L}\mathcal{F}^{0,0}, \label{rhot}\\
  \frac{d}{dt}\mathcal{F}^{1,0}&=&\mathcal{L}\mathcal{F}^{1,0}+(\mathcal{J}_{L+}-\mathcal{J}_{L-})\mathcal{F}^{0,0},\\
  \frac{d}{dt}\mathcal{F}^{0,1}&=&\mathcal{L}\mathcal{F}^{0,1}+(\mathcal{J}_{R+}-\mathcal{J}_{R-})\mathcal{F}^{0,0},\\
  \frac{d}{dt}\mathcal{F}^{2,0}&=&\mathcal{L}\mathcal{F}^{2,0}+2(\mathcal{J}_{L+}-\mathcal{J}_{L-})\mathcal{F}^{1,0}
  +(\mathcal{J}_{L+}+\mathcal{J}_{L-})\mathcal{F}^{0,0},\\
  \frac{d}{dt}\mathcal{F}^{0,2}&=&\mathcal{L}\mathcal{F}^{0,2}+2(\mathcal{J}_{R+}-\mathcal{J}_{R-})\mathcal{F}^{0,1}
  +(\mathcal{J}_{R+}+\mathcal{J}_{R-})\mathcal{F}^{0,0},\\
  \frac{d}{dt}\mathcal{F}^{1,1}&=&\mathcal{L}\mathcal{F}^{1,1}+(\mathcal{J}_{L+}-\mathcal{J}_{L-})\mathcal{F}^{0,1}
  +(\mathcal{J}_{R+}-\mathcal{J}_{R-})\mathcal{F}^{1,0}, \quad\mbox{etc.}
\end{eqnarray}
As described above, these coefficients contain the full counting
statistics. The charge
current out of lead $l$ is defined as the negative of the time-derivative of
the charge in lead $l$, $I_l(t)=e\,dN_l/dt$. The final
expression for the current out of the left lead is given
by\cite{AP16702}
\begin{eqnarray}\label{I2}
\langle
I_L(t)\rangle&=&e\,\mathrm{Tr_{dot}}\,\langle\dot\mathcal{F}^{1,0}\rangle
=e\,\mathrm{Tr_{dot}}
  \, (\mathcal{J}_{L+}-\mathcal{J}_{L-})\,\mathcal{F}^{0,0} .
\end{eqnarray}
The dc component of the current then gives the time average
$\bar I$. In order to find $\mathcal{F}^{0,0}$, we have to solve a
set of linear equations with the help of the normalization condition
of probability $\mathrm{Tr_{dot}}\, \mathcal{F}^{0,0}(t) =1$.

\subsubsection{Generalized MacDonald formula for time-averaged noise spectra}

We are interested in the frequency-dependent current noise of the
quantum dot driven by an ac field. The zero-frequency current
noise for non-adiabatical driving has been investigated in
Ref.\ \onlinecite{AP16702} using the Floquet master-equation approach
in the Coulomb-blockade regime. The symmetrized current-current correlation
function is defined by
\begin{eqnarray}
  S_{ll'}(t,t')=\langle \hat I_l(t)\hat I_{l'}(t')\rangle+\langle
  \hat I_{l'}(t')\hat I_l(t)\rangle -2\langle \hat I_l(t)\rangle\langle
  \hat I_{l'}(t')\rangle,
\label{S.tt}
\end{eqnarray}
where $\hat I_l(t)$ represents the current operator at the time $t$
from the lead $l$. The current-noise spectra are defined as the
Fourier transform of $S_{ll'}(t,t')$. Since our system is driven by
an ac field, the current noise is a double-time function.  However,
the periodicity of our problem makes it possible to characterize the
spectra by averaging over one driving period.

 At finite frequencies, the total current $I(t)$ measured by a
measurement device depends on both the particle and the displacement currents in
the lead-dot-lead junction. If one expresses the displacement currents by the
particle currents $I_L$ and $I_R$, one obtains the Ramo-Shockley
theorem,\cite{Ramo1939,Shockley1938,PR3361} 
\begin{eqnarray}
  I(t)=a I_L(t)-b I_R(t).
\end{eqnarray}
Here the coefficients $a$ and $b$, which satisfy
$a+b=1$, are specified by the device geometry. It is straightforward
to show that the total time-averaged noise spectrum is given by
\begin{eqnarray}
  \bar S(\omega)=a^2\bar S_{LL}(\omega)+b^2\bar S_{RR}(\omega)-ab(\bar S_{LR}(\omega)+\bar
  S_{RL}(\omega)),
\end{eqnarray}
where  $\bar S_{ll'}(\omega)$ ($l,l'=L,R$) represents the
frequency-dependent time-averaged current correlation between $I_l$
and $I_{l'}$,
\begin{eqnarray}
  \bar S_{ll'}(\omega)=\frac{1}{\mathcal{T}}\int^\mathcal{T}_0
  dt\int dt'\, e^{i\omega (t-t')}
  S_{ll'}(t,t').
\end{eqnarray}
In this study, we used two counting fields to derive the noise
spectra. An alternative approach is to calculate the charge fluctuation on the
dot employing the quantum regression
formula.\cite{MasterEquation,QuantumNoise,PRL92206601,PRB70205334}
The two approaches are physically equivalent due to the charge
conservation condition in the transport.

The formula for the zero-frequency noise has been presented in Ref.\
\onlinecite{AP16702}. For the two-terminal device, it is adequate to
find the time-averaged zero-frequency noise from the fluctuations of
the current flowing out of a chosen lead, $\bar S(0)=\bar
S_{ll}(0)$. The solution for $\bar S(0)$ resulting from the Floquet
quantum master equation reads
\begin{eqnarray}
  \bar S(0)=\frac{2}{\mathcal{T}}\int^\mathcal{T}_0dt\,
e^2\,\mathrm{Tr_{dot}}\left[2(\mathcal{J}_{l+}-\mathcal{J}_{l-})\mathcal{F}^{
\delta_{lL},\delta_{lR}}_{\perp}
  +(\mathcal{J}_{l+}+\mathcal{J}_{l-})\mathcal{F}^{0,0}\right],
\end{eqnarray}
where the prefactor of $2$ is inserted to make the noise formula
consistent with Ref.\ \onlinecite{PR3361}. For Poissonian noise, we
then obtain $\bar S(0)=2e\bar I$. $\delta_{ij}$ is the usual
Kronecker symbol. Following Ref.\ \onlinecite{AP16702}, the new
function $\mathcal{F}^{\delta_{lL},\delta_{lR}}_{\perp}$ in the
noise expression is defined  as
\begin{eqnarray}
  \mathcal{F}^{\delta_{lL},\delta_{lR}}_{\perp}
  =\mathcal{F}^{\delta_{lL},\delta_{lR}}-\mathcal{F}^{0,0}\,\mathrm{Tr_{dot}}
  \{\mathcal{F}^{\delta_{lL},\delta_{lR}}\},
\end{eqnarray}
and satisfies the equation of motion
\begin{eqnarray}
  \dot{\mathcal{F}}^{\delta_{lL},\delta_{lR}}_{\perp}
  =\mathcal{L}(t)\mathcal{F}^{\delta_{lL},\delta_{lR}}_{\perp}
  +\left(\mathcal{J}_{l+}-\mathcal{J}_{l-}-\frac{1}{e}\langle
  I_l(t)\rangle\right)\mathcal{F}^{0,0}.
\end{eqnarray}
An efficient method to find the noise spectrum is
provided by the MacDonald formula,\cite{RPP1256} which has
been widely used in quantum
transport.\cite{PRB74075328,PRB75045340,PRB76085325,JAP104033532}
The validity of this formula requires that the current correlation
function $\langle I(t_1)I(t_2)\rangle$ is only a function of the
time difference $t_1-t_2$ and that, therefore, the transport is in the
stationary regime. A direct application of the MacDonald formula to
the present time-dependent transport problem is thus not possible.
However, in the present
study the driving field is time-periodic. The discrete temporal translation
symmetry $H(t+\mathcal{T})=H(t)$ makes it possible to generalize
the MacDonald formula for the noise spectra time-averaged
over one period as
\begin{eqnarray}
\frac{\bar S_{ll'}(\omega)}{\omega}
  &=&\frac{2e^2}{\mathcal{T}}\int^{\mathcal{T}}_0
  \frac{dt}{2i}\,\mathrm{Tr_{dot}}\left[\mathcal{S}(-i\omega)
  \mathcal{F}^{\delta_{lL}+\delta_{l'L},\delta_{lR}+\delta_{l'R}}(-i\omega)
-\mathcal{S}(i\omega)\mathcal{F}^{\delta_{lL}+\delta_{l'L},\delta_{lR}+
\delta_{l'R}}(i\omega)\right]  ,\qquad \label{MacDonald}
\end{eqnarray}
where the composents of the superoperator $\mathcal{S}(s)$ are given by
\begin{eqnarray}
[\mathcal{S}(s)]_{lmk;l'm'k'}=(s-ik\Omega)\,\delta_{ll'}\delta_{mm'}
  \delta_{kk'}.
\end{eqnarray}
To clarify the meaning of this definition, we note that the
superoperator $\mathcal{S}(s)$ acts on an arbitrary operator $A$
with Fourier-transformed matrix elements $A_{lm;k}$ in our standard
basis $|l\rangle, |m\rangle = |0\rangle,|\uparrow\rangle,
|\downarrow\rangle, |\uparrow\downarrow\rangle$ as
\begin{eqnarray}
[\mathcal{S}(s)\,A]_{lm;k} = \sum_{l'm'}\sum_{k'}
  [\mathcal{S}(s)]_{lmk;l'm'k'}\, A_{l'm';k'} .
\end{eqnarray}
The derivation of the generalized MacDonald formula is outlined in the
appendix. In evaluating the current noise from the generalized
MacDonald formula, we encounter the Laplace transforms $\mathcal{F}^{m,n}(s)$
of the moments of the electron
number operators in the leads. These moments are nothing but the expansion
coefficients of $\mathcal{F}$ defined in Eq.\ (\ref{Fmn}).

Taking the trace over the dot degrees of freedom and the average
over one period makes only the matrix elements
$\mathcal{F}^{m,n}_{ll;k=0}$ of $\mathcal{F}^{m,n}$ that are
diagonal in the dot basis and have Floquet index $k=0$ contribute to
the final result. We arrive at the expression
\begin{eqnarray}\label{MD}
  \frac{\bar
S_{ll'}(\omega)}{\omega}&=&-e^2\omega\, \mathrm{Tr_{dot}}
  [\mathcal{F}^{\delta_{lL}+\delta_{l'L},\delta_{lR}+\delta_{l'R}}_{k=0}(-i\omega)
  +\mathcal{F}^{\delta_{lL}+\delta_{l'L},\delta_{lR}+\delta_{l'R}}_{k=0}(i\omega)].
\end{eqnarray}
 Now we require the charge moments in the left and right leads.
They can be obtained from the equation of motion for $\mathcal{F}$
[Eq.\ (\ref{EOMF})]. Suppose we switch on the counting fields at
some time $t_1$, before that time the system can be described by the
density matrix without the counting fields in the (quasi-)
stationary limit. Since we have assumed $t_0\to -\infty$ above, any
initial correlation have died out at time $t_1$.\cite{PRL100150601}
We set the number of electrons having tunneled into lead $l$ up to
time $t_1$ to zero, $N_l(t_1,t_1)$=0. After time $t_1$, the system
evolves under the influence of the counting fields. Then, we solve
the equations for $\mathcal{F}^{m,n}(t)$ by means of Laplace
transformation. For example, the solution of Eq.\ (\ref{rhot}) reads
\begin{eqnarray}
\mathcal{F}^{0,0}(s)=[\mathcal{S}(s)-\mathcal{L}]^{-1}\,\mathcal{F}^{0,0}(t_1),
\end{eqnarray}
where $\mathcal{F}^{0,0}(t_1)$ can be found from the stationary
master equation $\mathcal{L}\mathcal{F}^{0,0}=0$ in the absence of
counting fields. Analogously, we find expressions for the other
expansion coefficients after the Laplace transformation as
\begin{eqnarray}
  \mathcal{F}^{1,0}=\left(\mathcal{S}-\mathcal{L}\right)^{-1}(\mathcal{J}_{1+}-\mathcal{J}_{1-})\mathcal{F}^{0,0},
\end{eqnarray}
\begin{eqnarray}
  \mathcal{F}^{0,1}=\left(\mathcal{S}-\mathcal{L}\right)^{-1}(\mathcal{J}_{2+}-\mathcal{J}_{2-})\mathcal{F}^{0,0},
\end{eqnarray}
\begin{eqnarray}
  \mathcal{F}^{2,0}=\left(\mathcal{S}-\mathcal{L}\right)^{-1}\left[2(\mathcal{J}_{1+}-\mathcal{J}_{1-})
  \mathcal{F}^{1,0}+(\mathcal{J}_{1+}+\mathcal{J}_{1-})\mathcal{F}^{0,0}\right],
\end{eqnarray}
\begin{eqnarray}
  \mathcal{F}^{0,2}=\left(\mathcal{S}-\mathcal{L}\right)^{-1}\left[2(\mathcal{J}_{2+}-\mathcal{J}_{2-})
  \mathcal{F}^{0,1}+(\mathcal{J}_{2+}+\mathcal{J}_{2-})\mathcal{F}^{0,0}\right],
\end{eqnarray}
\begin{eqnarray}
  \mathcal{F}^{1,1}=\left(\mathcal{S}-\mathcal{L}\right)^{-1}\left[(\mathcal{J}_{1+}-\mathcal{J}_{1-})
  \mathcal{F}^{0,1}+(\mathcal{J}_{2+}-\mathcal{J}_{2-})\mathcal{F}^{1,0}\right],
\end{eqnarray}
where we have omitted the arguments $s$.
The solutions for these coefficients together with the generalized
MacDonald formula [Eq.\ (\ref{MacDonald})] give the desired
time-averaged current-noise spectra of the ac-driven quantum dot.
The approach
presented in this study can easily be generalized to take
more complex structures with multiple levels and
inter-level transitions into account.

\section{Results and discussion}
\label{sec:results}

In the following, we present our numerical results based on the
Floquet master-equation method and discuss the transport properties
of the single-level quantum dot with time-dependent fields.
An additional dc magnetic field along the $z$ or
$x$ direction is taken into account; it splits the energy levels of
the singly charged quantum dot due to
the Zeeman effect. In the present study, we choose the ac field to
be either a rotating magnetic field in the $xy$ plane or an ac
gate voltage. The ac gate voltage
and the rotating magnetic field will affect the quantum conductor in
quite different manners. An ac gate voltage only
changes the eigenvalues of $H_\mathrm{dot}(t)$ periodically, which in the
adiabatic limit of large $\mathcal{T}$ become the eigenenergies.
The ac gate voltage will not induce any transition between different
eigenstates of $H_\mathrm{dot}(t)$ (no spin flip is possible)
because the eigen\emph{states} are unaffected by the gate voltage. The
electron is trapped in one spin state. The situation
is different for a rotating magnetic field. A rotating
magnetic field does not change the eigenvalues of $H_\mathrm{dot}(t)$ but
does change the eigenstates and thus can flip the
spin of the electron. The spin polarization of the dot
will thus evolve with the rotating magnetic field. The two types of
ac fields show drastically different
behaviors in the transport properties as we will show below.

The full Hamiltonian of the quantum dot is written as (we reiterate
that we choose $|e|=\hbar=k_B=1$),
\begin{eqnarray}
  H_{\mathrm{dot}}&=&\sum_\sigma(eV_G+V_{\mathrm{ac}}\cos \Omega
  t +\sigma B_z)\, d^\dag_\sigma d_\sigma +U\, d^\dag_\uparrow d_\uparrow
  d^\dag_\downarrow d_\downarrow \nonumber\\
&&{}+B_x\, (d^\dag_\uparrow
  d_\downarrow +d^\dag_\downarrow d_\uparrow) +
  B_{\mathrm{ac}}\,(d^\dag_\uparrow
  d_\downarrow e^{i\Omega t}+d^\dag_\downarrow d_\uparrow
  e^{-i\Omega t}),
\label{full.Hdot}
\end{eqnarray}
where $eV_G$ is the on-site energy of the quantum dot due to the dc
component of the gate voltage $V_G$, $V_{\mathrm{ac}}$ is the
amplitude of the oscillating gate voltage, $U$ represents the
intra-dot Coulomb interaction, and $B_x$ and $B_z$ are half the
Zeeman energies of the singly occupied dot due to the dc magnetic
fields in the $x$ and $z$ direction, respectively. Half the Zeeman
energy of the rotating magnetic field is given by $B_{\mathrm{ac}}$.
Note that while it is customary to talk about photon-assisted
processes in this context, the treatment of the electromagnetic
field in the Hamiltonian is completely classical.

We work in the sequential-tunneling regime and choose a symmetric
coupling geometry with $a=b$. We assume that the bias voltage
$V_\mathrm{dc}$ symmetrically shifts the chemical potentials by
$\mu_{L,R}=\pm eV_\mathrm{dc}/2$. In the framework of wide-band
approximation, the tunneling rate is given by
$\Gamma^{l\sigma}_{mn;ij}(\epsilon)={2\pi}{}V^{ij}_{l,k,\sigma}V^{mn*}_{l,k,\sigma},$
where we have assumed the coupling strength $V_{l,k,\sigma}^{ij}=V$
to be a constant and have set the density of states of lead $l$ to
unity. In the present study, we have assumed the tunneling matrix to
be independent of the energy and the occupation number on the dot.
An inclusion of state-dependent tunneling is straightforward. We
assume that the electrons tunneling in and out of the dot with an
energy-independent rates $\Gamma=\Gamma^{l\sigma}_{mn;ij}(\epsilon)$
and set $\Gamma=1$ as the energy unit.

\subsection{Differential conductance}

\begin{figure}
\includegraphics[angle=0,width=0.8\textwidth]{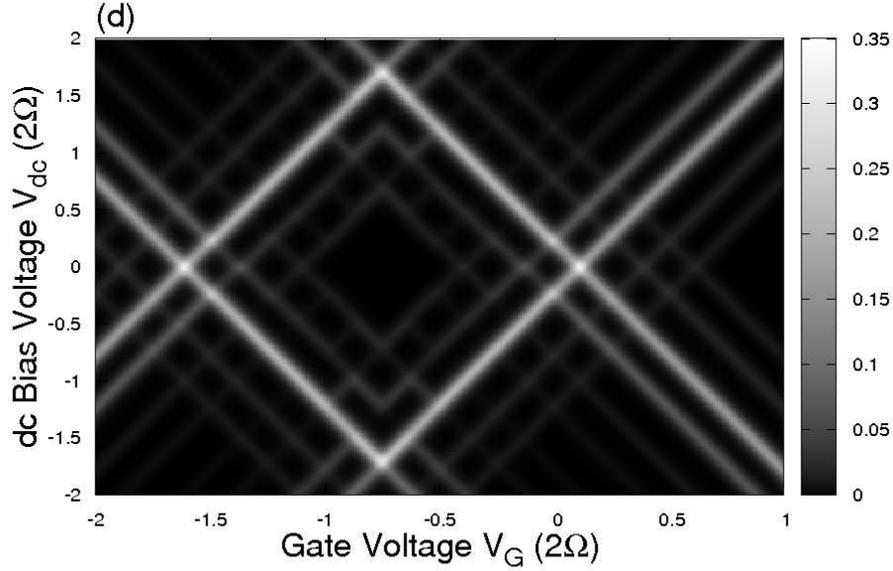} 
\caption{Gray-scale plots of the differential conductance
$dI/dV_\mathrm{dc}$ as a function of the dc bias voltage and the dc
gate voltage for a quantum dot (a) without any ac fields, (b)
modulated by an ac gate voltage, (c) modulated by a rotating
magnetic field, and (d) modulated by a rotating magnetic field and
with an additional dc magnetic field applied in the plane of the
rotating field. Dark regions represent low differential conductance.
The frequency of the ac field is $\Omega=8$. The amplitude of the ac
gate voltage in (b) and of the rotating magnetic field in (c), (d)
are $V_\mathrm{ac}=6.4$ and $B_\mathrm{ac}=3.2$, respectively. The
dc magnetic field in (d) is $B_x=1.6$.} \label{dIdV}
\end{figure}

We start our discussion with the differential conductance. The
gray-scale plot Fig.\ \ref{dIdV} shows the differential conductance
$dI/dV_\mathrm{dc}$ vs.\ the dc bias voltage $V_\mathrm{dc}$ and the
gate voltage $V_G$ with or without an ac field. The calculations are
for the Coulomb interaction strength $U=24$ and the temperature
$k_BT=0.32$. The frequency of the ac field is $\Omega=8$.

Without an ac field, Fig.\ \ref{dIdV}(a) gives the familiar diamond
structure due to the Coulomb blockade. Numerical results for the
differential conductance when the quantum dot is modulated by an ac
gate voltage are presented in Fig.\ \ref{dIdV}(b). Fig.\ \ref{dIdV}(c)
gives the results when the quantum dot is modulated by a
rotating magnetic field. Fig.\ \ref{dIdV}(d) shows the differential
conductance when the quantum dot is modulated by a rotating
magnetic field in the $xy$ plane while a dc magnetic field is applied
in the
$x$ direction, i.e., in the plane of the rotating magnetic field.

When there is an ac field, several striking features emerge in the
differential conductance: (1) At the edge of the Coulomb diamond,
the sharp differential-conductance peak for the dc transport shown
in Fig.\ \ref{dIdV}(a) is partially suppressed by the ac gate
voltage or the rotating magnetic field. Note the different gray
scales in Fig.\ 1 (a), (b), (c), and (d). This can be attributed to
the suppression of the elastic resonant peak by the photon-assisted
processes. (2) In the presence of an ac field, there are lines
parallel to the edges of the Coulomb diamond. The distance of these
lines to the peak position is approximately the frequency of the ac
field, indicating a photon-assisted tunneling process. (3) An
interesting feature of these lines can be observed inside the
Coulomb diamonds. For the ac gate voltage, the Floquet quasienergies
are spin degenerate. Therefore, the main lines in the differential
conductance plot in Fig.\ \ref{dIdV}(b) are not split. However,
satellites due to photon-assisted inelastic tunneling events appear,
in which an energy quantum of $\Omega$ is absorbed from or emitted
into the driving field. We see from Fig.\ \ref{dIdV}(b) that these
additional lines remain distinct inside the Coulomb diamond. On the
other hand, when the quantum dot is driven by a rotating magnetic
field, the quasienergies are not degenerate. Therefore, the main
elastic lines are split into two at the edge of the Coulomb diamond
in Fig.\ \ref{dIdV}(c). Interestingly, the lines due to the
photon-assisted tunneling now only appear outside of the Coulomb
diamond, as can be seen in Fig.\ \ref{dIdV}(c), indicating that the
photon-assisted tunneling is forbidden inside the Coulomb diamond.
When the quantum dot is modulated by a rotating magnetic field and a
dc magnetic field is applied in the plane of the rotating field,
these lines in the Coulomb-blockade regime revive. This can be
clearly seen in Fig.\ \ref{dIdV}(d).

The disappearance of the photon-assisted tunneling inside the
Coulomb diamond for a pure rotating magnetic field can be understood
as follows. In the Coulomb diamond, the Floquet quasienergies
corresponding to the singly occupied states are far below the Fermi
energies of the two leads. A direct tunneling between the dot and
the leads is forbidden due to the Pauli principle and Coulomb
blockade. Therefore, an electron is effectively trapped in one
quantum state on the dot. According to our previous discussion, only
the spin direction of this quantum state can evolve with the
rotating magnetic field. However, its eigenvalues of
$H_\mathrm{dot}(t)$ remain unchanged. Therefore, the electron cannot
gain extra energy from the ac magnetic field. As a consequence, we
cannot observe lines due to photon-assisted tunneling inside the
Coulomb diamond. Outside of the Coulomb diamond, the tunneling
between the dot and the leads becomes possible. When an electron is
injected from the lead into the dot, the system can absorb or emit
photons, i.e., the Floquet index $k$ can change. One could say that transport
happens via several Floquet channels. Such
photon-mediated tunneling can then give rise to the photon-assisted
differential-conductance peaks.

The situation becomes different when a dc magnetic field is applied
in the plane of the rotating magnetic field as shown in Fig.\
\ref{dIdV}(d). In that case, the eigenstates and the eigenvalues of
$H_\mathrm{dot}(t)$ change periodically. Electrons can gain extra
energy from the ac field by absorbing or emitting a photon. In the
Coulomb diamond, electrons on the dot are able to tunnel out via the
photon-assisted tunneling and we again find the lines due to the
photon-assisted differential-conductance peaks inside the Coulomb
diamond as shown in Fig.\ \ref{dIdV}(d).

\subsection{Zero-frequency Fano factor}

\begin{figure}
\includegraphics[width=0.8\textwidth]{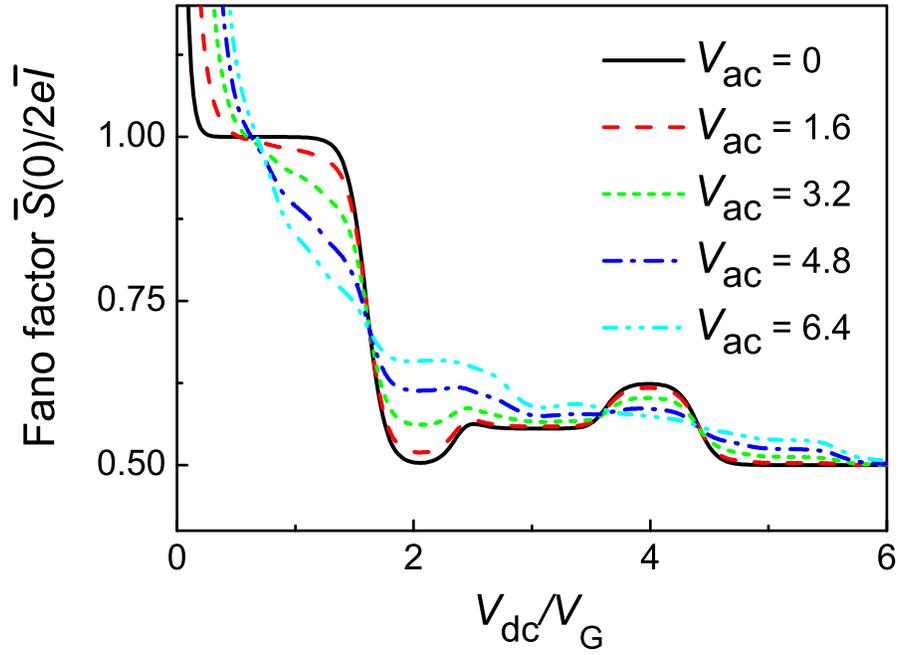}
\caption{(Color online) Fano factor of the quantum dot as a function
of dc bias voltage $V_{dc}$ in unit of $V_G$ for different
amplitudes $V_\mathrm{ac}$ of the ac gate voltage. The parameters of
the device are given in the text.}\label{FanoVac}
\end{figure}

In the following, we show numerical results for the time-averaged
zero-frequency noise of the quantum dot. The zero-frequency noise
has been studied by the quantum master-equation method in the
stationary \cite{PR3361,PRB68115105,PRB76085325} and also in the
time-dependent case.\cite{AP16702} Without ac field and at zero
temperature, the zero-frequency Fano factor $S(0)/2eI$ describes the
deviation of the shot noise from its Poissonian value. We choose the
parameters $T=0.32$, $U=8$, $eV_G=8$, and $\Omega=4.8$. We assume
that a dc magnetic field in the $z$ direction, $B_z=1.6$, is applied
to the quantum dot. The finite value of $U$ and the Zeeman splitting
make it possible to see plateaus in the Fano factor at different
occupation numbers on the dot.\cite{PRB68115105} Fig.\ \ref{FanoVac}
shows the zero-frequency Fano factor as a function of the dc bias
with or without an ac gate voltage.

Without an ac gate voltage, $V_\mathrm{ac}=0$, the results reproduce
the main features reported in Ref.\ \onlinecite{PRB68115105}. At
very low bias voltage $V_\mathrm{dc}\rightarrow 0$, the main
contribution to the noise is the finite thermal noise while the
current as well as the shot noise are suppressed. Therefore, the
Fano factor diverges at $V_\mathrm{dc}=0$. For low dc bias voltage,
the quantum dot operates in the Coulomb-blockade regime. With
further increasing dc bias voltage, the energy levels of the quantum
dot one by one enter the transport window defined by the dc bias.
This can be clearly identified in the Fano factor by the plateaus at
different values. The edges of the plateaus are broadened by the
finite temperature. For very large dc bias, where all the energy
levels of the quantum dot lie in the transport window, the Fano
factor approaches the well-known limit of $1/2$ for our
symmetric-coupling case.\cite{PRB68115105}

\begin{figure}
\includegraphics[width=0.8\textwidth]{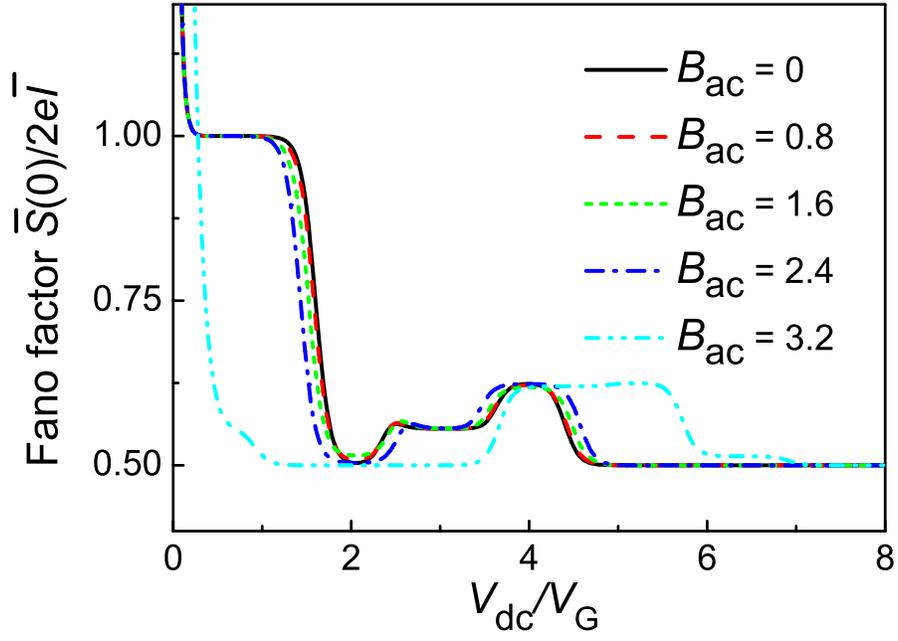}
\caption{(Color online) Fano factor of the quantum dot as a function
of dc bias voltage $V_{dc}$ in unit of $V_G$ for different
amplitudes $B_\mathrm{ac}$ of the rotating magnetic field. The other
parameters are the same with those of Fig.
\ref{FanoVac}.}\label{FanoRac}
\end{figure}

The results for the dc case
demonstrate that the plateaus of the Fano factor can
give a good account of the transport channels.\cite{PRB68115105}
In the presence of an ac field, we now consider the time-averaged Fano
factor $\bar{S}(0)/2e\bar{I}$.
We can see
from Fig.\ \ref{FanoVac} that for small $V_\mathrm{dc}$
the Fano
factor becomes larger as we increase the amplitude of the ac gate
voltage.  On the other hand,  additional photon-assisted
transport channels are available due to the  ac field, which will modify
the Fano-factor curve.
With increasing ac field, the Fano factor will thus
deviate from the plateau behavior seen in dc case due to the opening
of these photon-assisted transport channels.
When the ac gate voltage is large enough, the Floquet eigenstates
that lie outside of the transport window can contribute to the
current via photon-assisted tunneling. As a consequence, the
plateaus in the Fano-factor curve become vague. For very large bias
voltages, all the Floquet levels are well inside the transport
window. The Fano factor then will approach the same value $1/2$ as
for the time-independent transport. In Fig.\ \ref{FanoRac}, we
present our results for the zero-frequency current noise in the
presence of a rotating magnetic field. As in the case of an ac gate
voltage, the Fano factor deviates from the dc behavior with
increasing ac field. Additional plateaus can be observed in the
Fano-factor curve when we vary the dc bias voltage. Transitions between
plateaus result from additional Floquet channels becoming available.

\subsection{Frequency-dependent Fano factor}

Now we present our results for the full current-noise spectra of a
quantum dot under an ac field. The noise spectra have previously
been studied in the stationary-transport regime. An analytical
expression for the noise spectrum of a single-level quantum dot can
be found in Ref.\ \onlinecite{PRB76085325}. Unless stated otherwise,
the following calculations assume $U=24$, $V_\mathrm{dc}=12.7$,
$T=1.6$, $\Omega=8$ and $eV_G=-8$. We introduce the
frequency-dependent Fano factor $\bar S(\omega)/2e\bar I$ to
characterize the time-averaged noise power. As discussed previously,
the ac gate voltage and the rotating magnetic field will modulate
the quantum conductor in different ways. In the following, we show
that the noise spectra are also strikingly different.

\begin{figure}
  \includegraphics[angle=0,width=0.8\textwidth]{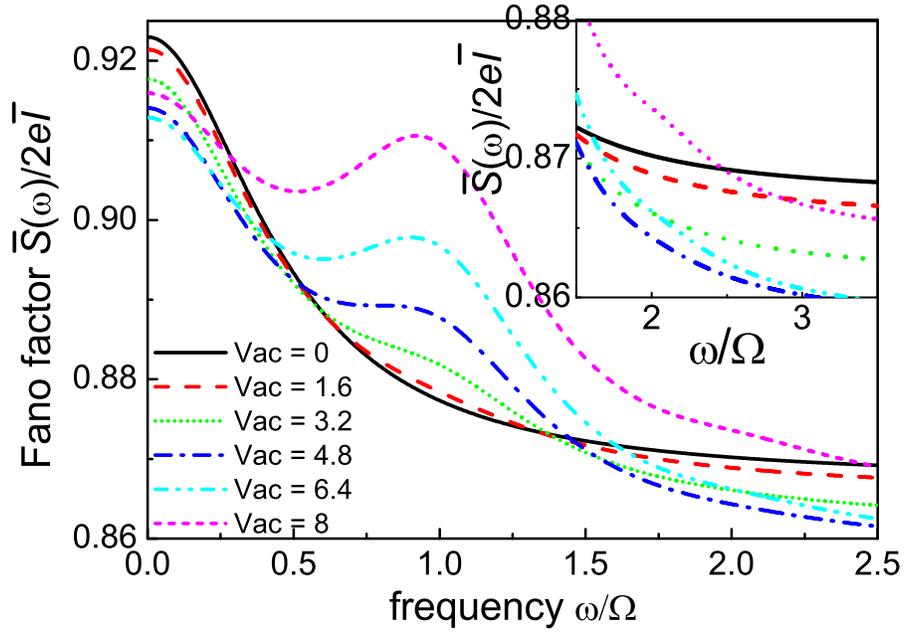}
  \caption{(Color online) Noise spectra for different gate voltage amplitudes
  $V_\mathrm{ac}$. Independently of $V_\mathrm{ac}$, the main peak position is fixed at $\Omega$.
  The inset shows an enlarged view of the noise spectra
  around $2\Omega$.}\label{SVac}
\end{figure}

In Fig.\ \ref{SVac}, we present the results for the
frequency-dependent Fano factor $\bar S(\omega)/2e\bar I$
as a function of the
frequency $\omega$ for different amplitudes $V_\mathrm{ac}$. No dc
magnetic field is applied. Without an ac field ($V_\mathrm{ac}=0$), the
noise spectrum shows a peak at zero frequency and approaches a
constant value for large $\omega$. The peak in the noise
spectrum is due to the elastic processes in the
transport.\cite{PRL102016802} When an ac gate voltage is applied,
additional structures in the noise spectra are expected due to
photon-assisted processes.  For the present set of
parameters, one can clearly see that with increasing amplitude of
the ac gate voltage, an additional peak appears in the noise
spectrum. While the height and width of this peak vary a lot
with increasing amplitude, its peak position $\omega_{p}$
remains almost unchanged at the external driving frequency $\Omega$.

\begin{figure}
  \includegraphics[angle=0,width=0.8\textwidth]{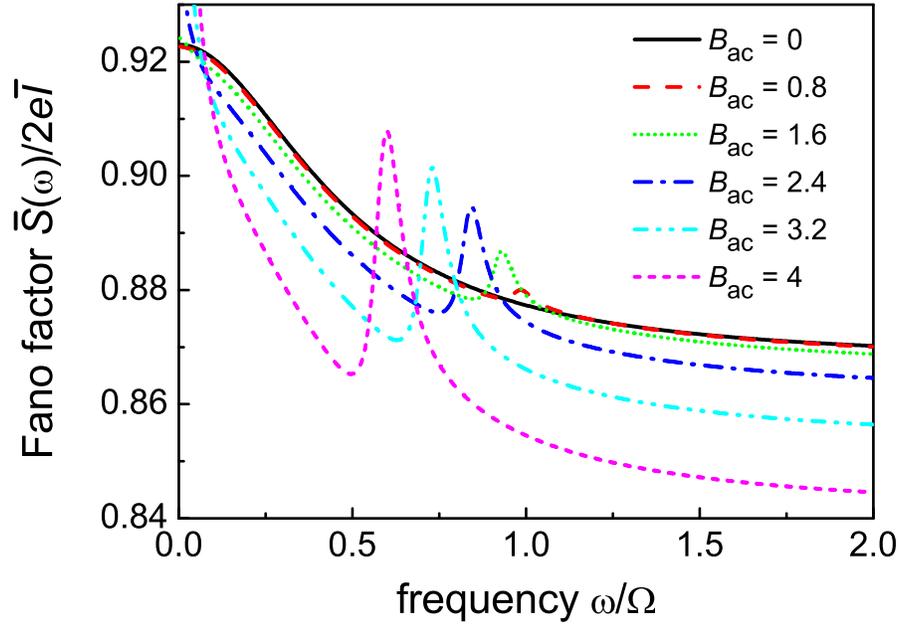}
  \caption{(Color online) Noise spectra with a rotating magnetic field in the
  $xy$ plane with various amplitudes $B_\mathrm{ac}$.
  The peak position shifts with increasing $B_\mathrm{ac}$.}\label{SRac}
\end{figure}

Now we turn to the rotating magnetic field in the $xy$ plane. In
Fig.\ \ref{SRac}, we plot the Fano factor  as a function of the
frequency $\omega$ for different amplitudes $B_\mathrm{ac}$ of the
rotating magnetic field. Similarly to the results presented in Fig.\
\ref{SVac}, a peak is generated and the width and height of this
peak depend on the amplitude. However, the peak position is not
fixed at $\Omega$ in contrast to what we have observed in Fig.\
\ref{SVac} for the ac gate voltage. Instead, its position shifts with
increasing amplitude, as shown in Fig.\ \ref{SRac}.

By comparing Fig.\ \ref{SVac} and Fig.\ \ref{SRac}, we see that the
peak position of the noise spectra behaves differently when we
increase the ac strength, depending on the type of the ac field.
Recalling that when electrons tunnel through a time-independent
quantum two level system, its current noise spectra show additional
structure at the energy difference of the two transport channels of
the system due to its internal coherent dynamics,\cite{PRL92206601}
we will show that the peak position of the noise spectra for ac
transport can be understood from the interference between two
possible Floquet transport channels. If the quantum dot is modulated
by a rotating magnetic field, the last term in the dot Hamiltonian
[Eq.\ (\ref{full.Hdot})] shows that the ac magnetic field couples
one spin state with the quasienergy $\epsilon$ with a state with the
opposite spin and the quasienergy $\epsilon-\Omega$ (in the extended
zone scheme). The coupling strength is given by $B_{\mathrm{ac}}$.
{The corresponding Floquet Hamiltonian then decomposes into $2\times
2$ blocks of the form
\begin{eqnarray}\label{hFl}
h_\mathrm{Fl} = \left(
    \begin{array}{cc}
      \epsilon & B_\mathrm{ac} \\
      B_\mathrm{ac} & \epsilon-\Omega \\
    \end{array}
  \right).
\end{eqnarray}
The resulting quasienergies in the first Brillouin zone $[0,\Omega)$ are
\begin{eqnarray}
\epsilon_1 &=&
  \epsilon-\frac{\Omega}{2}+\frac{\sqrt{\Omega^2+4B_\mathrm{ac}^2}}{2} , \\
\epsilon_2 &=&
  \epsilon+\frac{3\Omega}{2}-\frac{\sqrt{\Omega^2+4B_\mathrm{ac}^2}}{2}
\end{eqnarray}
with the difference
\begin{eqnarray}\label{EqPeak}
\omega_p=\epsilon_2-\epsilon_1=2\Omega-\sqrt{\Omega^2+4B_\mathrm{ac}^2}
\end{eqnarray}
(these expressions hold if $B_\mathrm{ac}<\sqrt{3}\,\Omega/2$).}

If now an electron tunnels into the dot, the system ends up in a
superposition of the two Floquet states, the phases of which change
with different angular frequencies, corresponding to spin precession
with the difference frequency $\omega_p$. When the electron tunnels
out again, the superposition is projected onto the spin direction of
the original electron since lead electron creation and annihilation
operators are paired with identical quantum numbers in the master
equation. This leads to interference with a typical frequency
$\omega_p$, which enhances the current-current correlation function
$S_{ij}(t,t')$ in Eq.\ (\ref{S.tt}) for $t-t'$ being a multiple of
the period $2\pi/\omega_p$ and thus leads to a peak in the noise
spectrum at $\omega_p$. The peaks seen in Fig.\ \ref{SRac} are
indeed centered at $\omega_p$ given by Eq.\ (\ref{EqPeak}).

Comparing with the stationary transport through a stationary two
level system,\cite{PRL92206601} the transport through a quantum dot
with rotating magnetic field can be understood as another type of
two level quantum system. The significant difference here is that
our two levels are defined by the Floquet channels due to a periodic
ac field and not by the true eigenenergies of an time-independent
Hamiltonian.

When an ac gate voltage is applied to the quantum dot, the ac field
will not couple the different spin states. Only the eigenvalues of
$H_\mathrm{dot}(t)$ will be modulated, see Eq.\ (\ref{full.Hdot}).
{The corresponding Floquet Hamiltonian decomposes into two infinite
blocks for the two spin directions, where each block has the
tridiagonal form}
\begin{eqnarray}
h_\mathrm{Fl} = \left(
    \begin{array}{ccccc}
      \ddots & & & & \\
      & \epsilon+\Omega & V_\mathrm{ac}/2 & 0 & \\
      & V_\mathrm{ac}/2 & \epsilon & V_\mathrm{ac}/2 & \\
      & 0 & V_\mathrm{ac}/2 & \epsilon-\Omega \\
      & & & & \ddots
    \end{array}
  \right).
\end{eqnarray}
Therefore, the electrons can tunnel through the quantum dot via
infinitely many Floquet channels with the same quasienergy in the
first Brillouin zone $[0,\Omega)$ but all possible Floquet indices
$k$. The quasienergies in the extended zone scheme thus differ by
integer multiples of $\Omega$. These quasienergy differences define
the peak positions in the noise spectra. In Fig.\ \ref{SVac}, a peak
at $\Omega$ appears, corresponding to two channels with their
Floquet indices (photon numbers) differing by unity. One should also
expect peak structures at $n\Omega$, $n>1$. However, to observe
these peak structures, one may need a stronger ac field to enable
multi-photon-assisted transport. In the inset of Fig.\ \ref{SVac}, a
small shoulder emerges at $2\Omega$ for the largest amplitude,
$V_\mathrm{ac}=8$.

\begin{figure}
  \includegraphics[angle=0,width=0.8\textwidth]{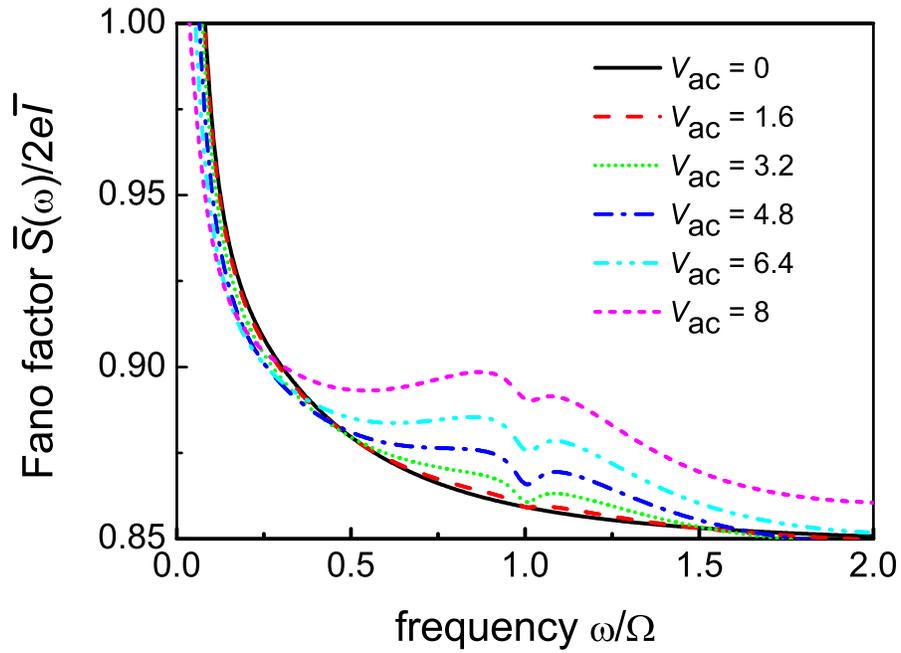}
  \caption{(Color online) Noise spectra for a quantum dot modulated by an
  oscillating gate voltage with various
  amplitudes $V_\mathrm{ac}$ in the presence of a dc magnetic field $B_z=1.6$
  in the $z$ direction. The dip position is fixed at the driving frequency.}\label{SVacZ}
\end{figure}

So far in our discussion, no dc magnetic field has been considered.
For the quantum dot with an ac gate voltage and a dc magnetic field
in the $z$ direction, our results of the noise spectra for different
voltage amplitudes are displayed in Fig.\ \ref{SVacZ}. The
parameters are the same as those used in Fig.\ \ref{SVac} except
that the strength of the dc magnetic field in the $z$ direction is
$B_z = 1.6$. For the present set of parameters, the peak at $\Omega$
is replaced by a dip. We observe that the appearance of the peak or
the dip depends on the detailed parameters used in our calculation.
The peak or dip position remains unchanged as we increase the
voltage amplitude. We have checked that the noise spectrum does not
depend on the direction of the dc magnetic field. This is because
the full SU(2) symmetry is preserved since we have included the time
evolution of the off-diagonal elements of the reduced density matrix
within our quantum master-equation approach.

\begin{figure}
  \includegraphics[angle=0,width=0.8\textwidth]{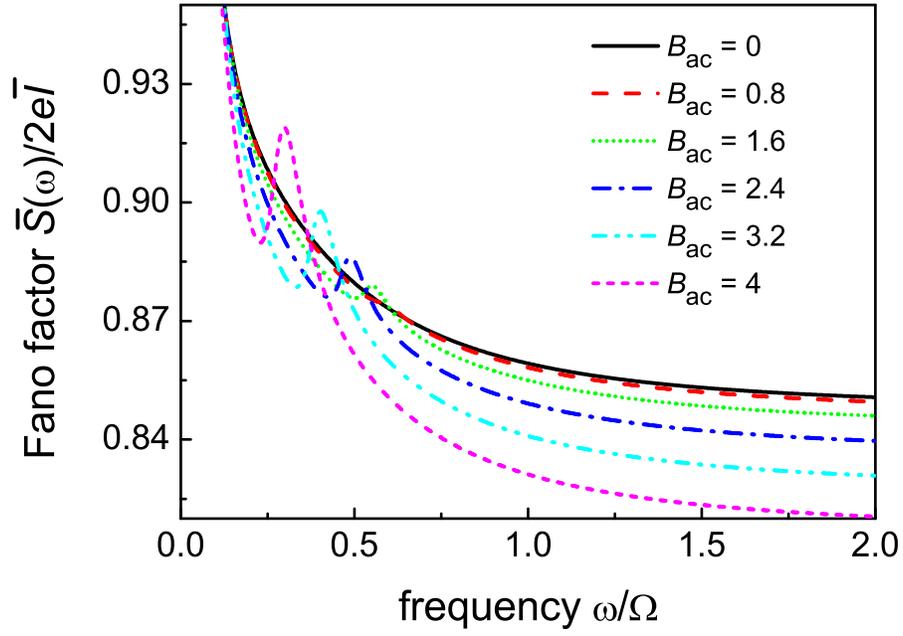}
  \caption{(Color online) Noise spectra in the presence of a rotating magnetic
  field in the $xy$ plane with various amplitudes $B_\mathrm{ac}$.
  A dc magnetic field $B_z=1.6$ is applied in the $z$ direction. Only one peak
  appears, which shifts with $B_\mathrm{ac}$.}\label{SRacZ}
\end{figure}

Contrary to the quantum dot with an ac gate voltage, for the case of
a rotating magnetic field, the noise spectra do depend on the
direction of the additional dc magnetic field. When the dc magnetic
field is perpendicular to the plane of the rotating magnetic field,
the noise spectra behave much like those for a pure rotating
magnetic  field. Only one peak appears at a non-zero frequency and
the peak position shifts with the amplitude of the rotating magnetic
field.  Numerical results for the noise with a rotating magnetic in
the $xy$ plane and a dc magnetic field in the $z$ direction are
displayed in Fig.\ \ref{SRacZ}. The parameters are the same as in
Fig.\ \ref{SRac} and the dc magnetic field is $B_{z}=1.6$. It is
easy to verify that the peak position can again be determined by the
difference between two Floquet quasienergies.

\begin{figure}
  \includegraphics[angle=0,width=0.8\textwidth]{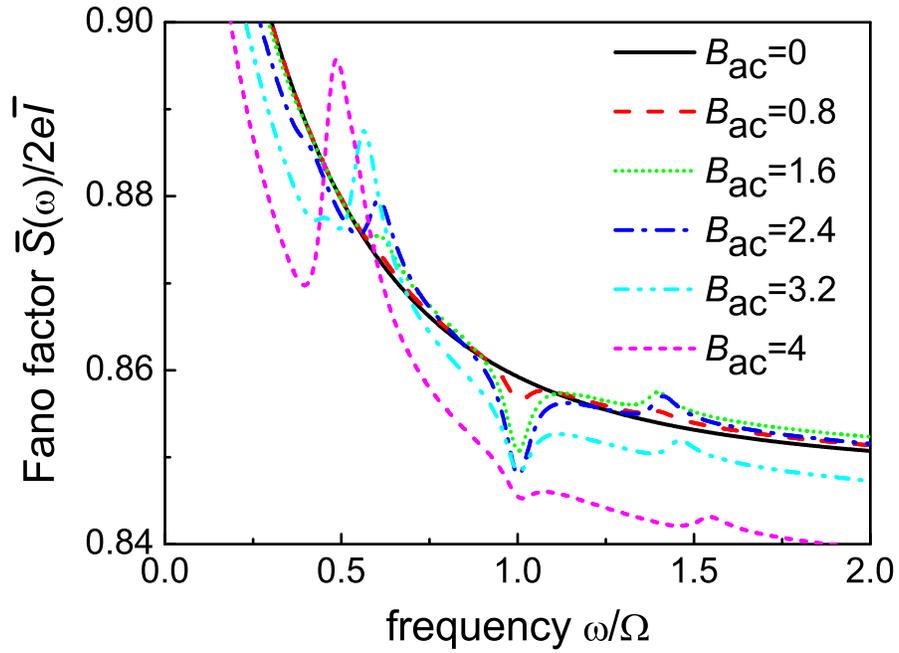}
    \caption{(Color online) Noise spectra with a rotating magnetic field in the
    $xy$ plane with various amplitudes $B_\mathrm{ac}$.
  A dc magnetic field $B_x=1.6$ is applied in the $x$ direction.
  New features appear in the noise spectra
  since more Floquet channels are involved in the transport in
  the presence of an in-plane magnetic field.
   }\label{SRacX}
\end{figure}

When a non-zero dc magnetic field $B_x$ is applied in the plane of
the rotating magnetic field, the Floquet Hamiltonian cannot be
reduced to a $2\times 2$ matrix form as for the previously discussed
situation of vanishing dc magnetic field, since the $B_x$ term in
Eq.\ (\ref{full.Hdot}) mixes spin-up and spin-down states. Together
with the rotating magnetic field this couples all Floquet states
with the same quasienergy in the first Brillouin zone and different
Floquet indices. Thus the electrons can tunnel through the quantum
dot via infinitely many Floquet channels. The interference between
these Floquet channels then gives rise to much richer behavior in
the noise spectra. Numerical results for the noise spectra of a
quantum dot driven by a rotating magnetic field in the $xy$ plane
and with a dc magnetic field in the $x$ direction are presented in
Fig.\ \ref{SRacX}. The parameters used in the calculation are the
same as in Fig.\ \ref{SRac} and the dc magnetic field in $B_x=1.6$.
In Fig.\ \ref{SRacX}, more peaks are observed than for vanishing dc
magnetic field in Fig.\ \ref{SRac}. One can see that besides the
peak determined by Eq.\ (\ref{EqPeak}), there are both peaks (dips)
fixed at $\Omega$ and structures at positions depending on the
ac-field amplitude $B_\mathrm{ac}$. All peak (dip) positions
correspond to the differences of available Floquet quasienergies.

\section{Summary}
\label{sec:sum}

In this paper, the transport properties of a single-level quantum
dot modulated by either an ac gate voltage or a rotating magnetic
field have been studied within the Floquet quantum master-equation
approach in the sequential-tunneling limit. We have
employed a generalized MacDonald formula  to obtain the
time-averaged current noise spectra
for both cases. Numerical results for the differential conductance
and the frequency-dependent current noise have been presented.
Besides the usual diamond structure due to the Coulomb blockade in
the differential conductance, photon-assisted tunneling can give
rise to additional lines parallel to the edges of the Coulomb
diamond. These lines cannot survive inside the Coulomb diamond in
the case of a rotating magnetic field. {This is due to the fact that
the rotating magnetic field  only periodically rotates the spin
direction while the energy of the electron on the dot remains
unchanged.} The frequency-dependent noise spectra of the quantum dot
show additional peaks or dips in the presence of an ac field. The
behavior of these additional structures depends on the nature of the
ac driving field. In the case of an ac gate voltage, the position of
the finite-frequency peak is fixed at the external ac frequency,
independently of the voltage amplitude. On the other hand, in the
case of a rotating magnetic field, the peak at non-zero frequency
moves with changing amplitude of the rotating magnetic field. An
additional dc magnetic field in the plane of the rotating magnetic
field can also drastically change the noise spectra: {it leads to
the appearance of both movable and fixed peak structures in the
noise spectra}. All these peak positions are found to be determined
by the energy differences between two Floquet transport channels.

\newpage

\appendix

\section{Derivation of the generalized MacDonald formula with ac field}
\label{app:B}

A derivation of the MacDonald formula for time-independent
transport has been presented in Ref.\ \onlinecite{PRB75045340}. We
show here that the periodicity of our time-dependent Hamiltonian makes it
possible to estimate the noise spectra by generalizing the MacDonald
formula. The derivation of the generalized MacDonald formula for the
time-averaged noise spectra is outlined in the following.

We start from the Fourier-transformed current correlation function
\begin{eqnarray}\label{AB1}
  S(t,\omega)=\int^\infty_{-\infty}d\tau\, e^{i\omega\tau}\left\langle\left\{
  \delta I(t),\delta I(t-\tau)\right\}\right\rangle,
\end{eqnarray}
where $\delta I(t)=I(t)-\langle I(t)\rangle$ and $\{A, B\}=AB+BA$. For
simplicity, we omit the lead index in the noise
expressions in this appendix.

From the definition of the current, we have
\begin{eqnarray}
  \int^{t+\tau}_t  dt'\, \delta I(t') &=&\int^{t+\tau}_t dt' \left[I(t')-\langle
  I(t')\rangle\right] \\ \nonumber
  &=&eN(t+\tau,t)-\int^{t+\tau}_t dt'\,\langle I(t')\rangle ,
\end{eqnarray}
where $N(t+\tau,t)$ denotes the number of charges transferred
during the interval from $t$ to $t+\tau$.
Taking the expectation value of the square of this equation, we obtain
\begin{eqnarray}\label{Eq4}
\lefteqn{
2e^2\left\langle\left[N(t+\tau,t)-\int^{t+\tau}_tdt'\langle
I(t')\rangle/e\right]\right\rangle
  ^2
  = \left\langle\int^{t+\tau}_tdt'\int^{t+\tau}_tdt''\left[\delta I(t')\delta I(t'')
  \delta I(t'')\delta I(t')\right]\right\rangle } \nonumber \\
&&  = \int^{t+\tau}_t dt'\int^{t+\tau}_tdt''\int^\infty_{-\infty}d\omega\,
  \frac{1}{2\pi} \, S(t',\omega)\,
  e^{i\omega(t'-t'')} .\hspace{15em}
\end{eqnarray}
Inserting the Fourier decomposition  of the time-dependent
relation
\begin{equation}
S(t',\omega)=S_0(\omega)+\sum_{k\neq 0}e^{-ik\Omega t'}
S_k(\omega)
\end{equation}
into Eq.\ (\ref{Eq4}), we obtain
\begin{eqnarray}
\ldots&=&\int^{t+\tau}_tdt'\int^\infty_{-\infty}d\omega\,\frac{1}{2\pi}\,
  \left[S_0(\omega)+\sum_k\nolimits'
e^{-ik\Omega t'}S_k(\omega)\right]e^{-i\omega t'}
\left[\frac{1}{i\omega}\,(e^{i\omega(t+\tau)} -e^{i\omega t})\right]\\
\nonumber &=&\int^\infty_{-\infty}d\omega\,\frac{1}{2\pi}\,S_0(\omega)\,
\frac{1}{\omega^2}\,(e^{-i\omega\tau}-1)(e^{i\omega\tau}-1)\\
\nonumber
&&+\int^\infty_{-\infty}d\omega\sum_k\nolimits'\frac{1}{2\pi}\,
  S_k(\omega)\,\frac{1}{\omega(\omega+k\Omega)}\,
  e^{-ik\Omega t}(e^{-i(\omega+k\Omega)\tau}-1)(e^{i\omega\tau}-1)
\\ \nonumber
&=&\int^\infty_{-\infty}d\omega\,\frac{1}{\pi}\,S_0(\omega)\,\frac{1}{\omega^2}\,
  (1-\cos(\omega\pi))\\
\nonumber
&&+\int^\infty_{-\infty}d\omega\sum_k\nolimits'\frac{1}{2\pi}\,S_k(\omega)\,
  \frac{e^{-ik\Omega
t}}{\omega(\omega+k\Omega)}(e^{-ik\Omega\tau}-e^{-i(\omega+k\Omega)\tau}-e^{i\omega\tau}+1)
,
\end{eqnarray}
where we have used the notation $\sum_k\nolimits'=\sum_{k\neq 0}$.
Differentiation with respect to $\tau$ gives
\begin{eqnarray}
\lefteqn{ \frac{d}{d\tau}\,2e^2\left\langle
\left[N(t+\tau,t)-\int^{t+\tau}_t dt'\, \langle I(t')\rangle/e
\right]^2\right\rangle
=\int^\infty_{-\infty}d\omega\frac{1}{\pi}\frac{S_0(\omega)}{\omega}\sin{\omega\tau}
 } \nonumber \\
&&+\int^\infty_{-\infty}d\omega\sum_k\nolimits'\frac{1}{2\pi}S_k(\omega)e^{-ik\Omega
t}\frac{1}{\omega(\omega+k\Omega)} \left(-ik\Omega e^{-ik\Omega
\tau}+i(\omega+k\Omega)e^{-i(\omega+k\Omega)\tau}-i\omega
e^{i\omega\tau}\right) . \qquad
\end{eqnarray}
We next perform a Fourier transformation
and take the time average over one period. The
second term on the right-hand side vanishes due to its periodicity.
Since the current correlation function is symmetric, $S(t,t')=S(t',t)$, it
can be shown that $\bar
S(\omega)=\frac{1}{\mathcal{T}}\int^\mathcal{T}_0 dt
\int^\infty_{-\infty} dt' S(t, t')e^{i\omega(t-t')}$ has the
property $\bar S(\omega)=\bar S(-\omega)$.
We arrive at the generalized formula for the time-averaged noise
spectrum for a periodic driving field,
\begin{eqnarray}
\lefteqn{ \frac{1}{\mathcal{T}}\int^\mathcal{T}_0dt\, 2e^2\int^\infty_{-\infty} d\tau\,
 e^{i\omega \tau} \frac{\partial}{\partial \tau}\left\langle
 [N(t+\tau,t)-\int^{t+\tau}_tdt'\langle I(t')\rangle/e ]^2\right\rangle } \nonumber \\
&& = \frac{1}{\mathcal{T}}\int^\mathcal{T}_0dt\, 2e^2\int^\infty_{-\infty}d\tau\,
 \sin(\omega \tau)\frac{\partial}{\partial \tau}\left\langle [N(t+\tau,t)-\int^{t+\tau}_tdt'\langle I(t')\rangle/e
 ]^2\right\rangle \nonumber \\
&& = 2i\,\frac{\bar S(\omega)}{\omega} .
\end{eqnarray}
 Noting that the integrand of the $\tau$ integral is even, we
obtain the final result for the generalized MacDonald formula for the
time-averaged noise spectrum,
\begin{eqnarray}\label{MDAppend}
  \frac{\bar S(\omega)}{\omega}=2e^2\,\frac{1}{\mathcal{T}}\int^\mathcal{T}_0 dt
  \int^\infty_{0}d\tau\,\sin \omega\tau\: \frac{\partial}{\partial
  \tau}\left\langle\left [N(t+\tau,t)-\int^{t+\tau}_tdt'\,
  \langle I(t')\rangle/e\right]^2\right\rangle.
\end{eqnarray}
In comparison to the MacDonald formula for steady state
transport, an
integration of $t$ over one period is carried out to obtain the
time-averaged noise spectra. The time average can also be
expressed by an average over the initial phase of the ac field. Hence, our
expression is equivalent to the form given by Clerk and
Girvin.\cite{PRB70121303}

\newpage

\bibliography{ref}

\newpage
\includegraphics[angle=0,width=0.8\textwidth]{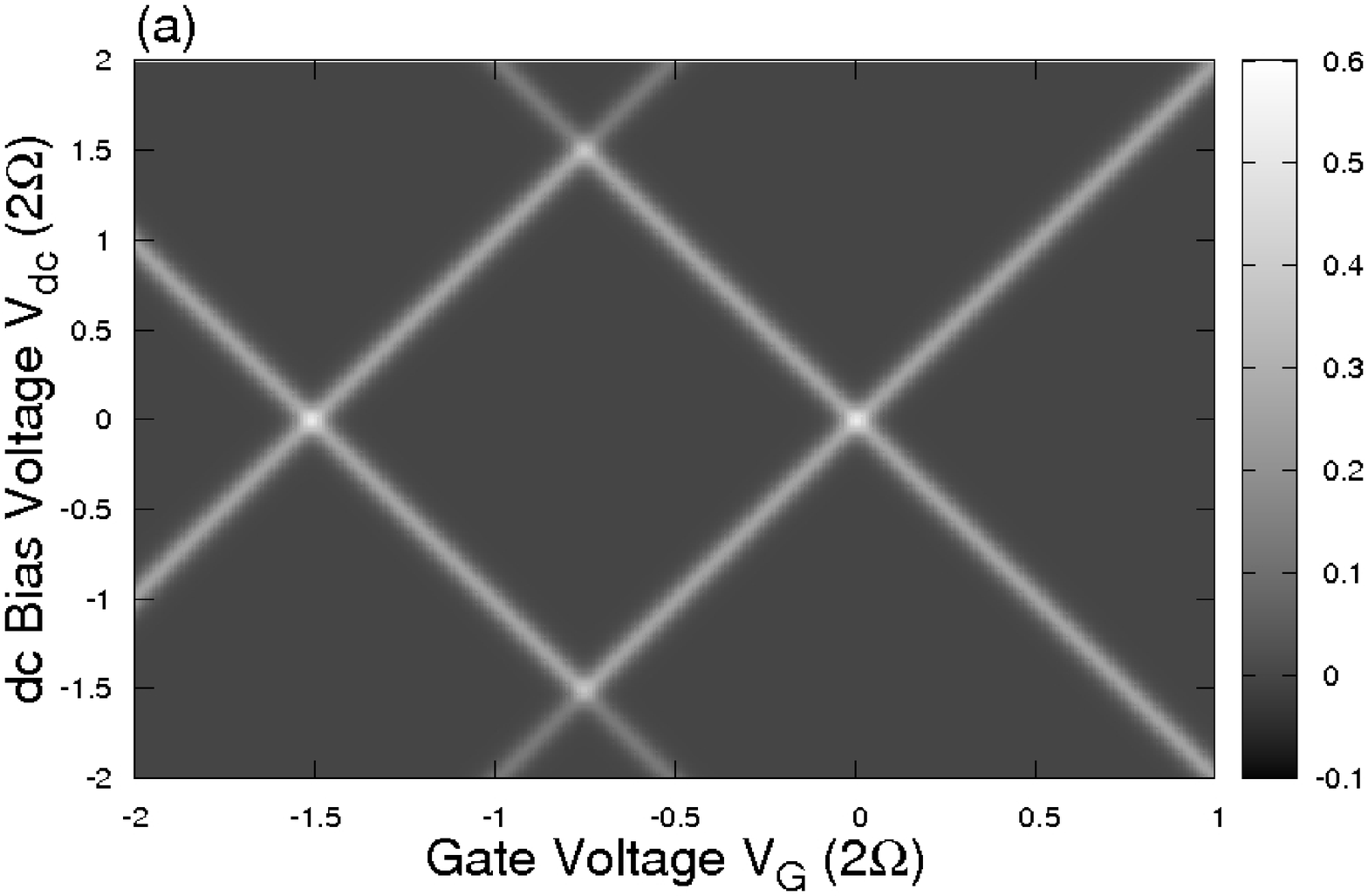} 


\includegraphics[angle=0,width=0.8\textwidth]{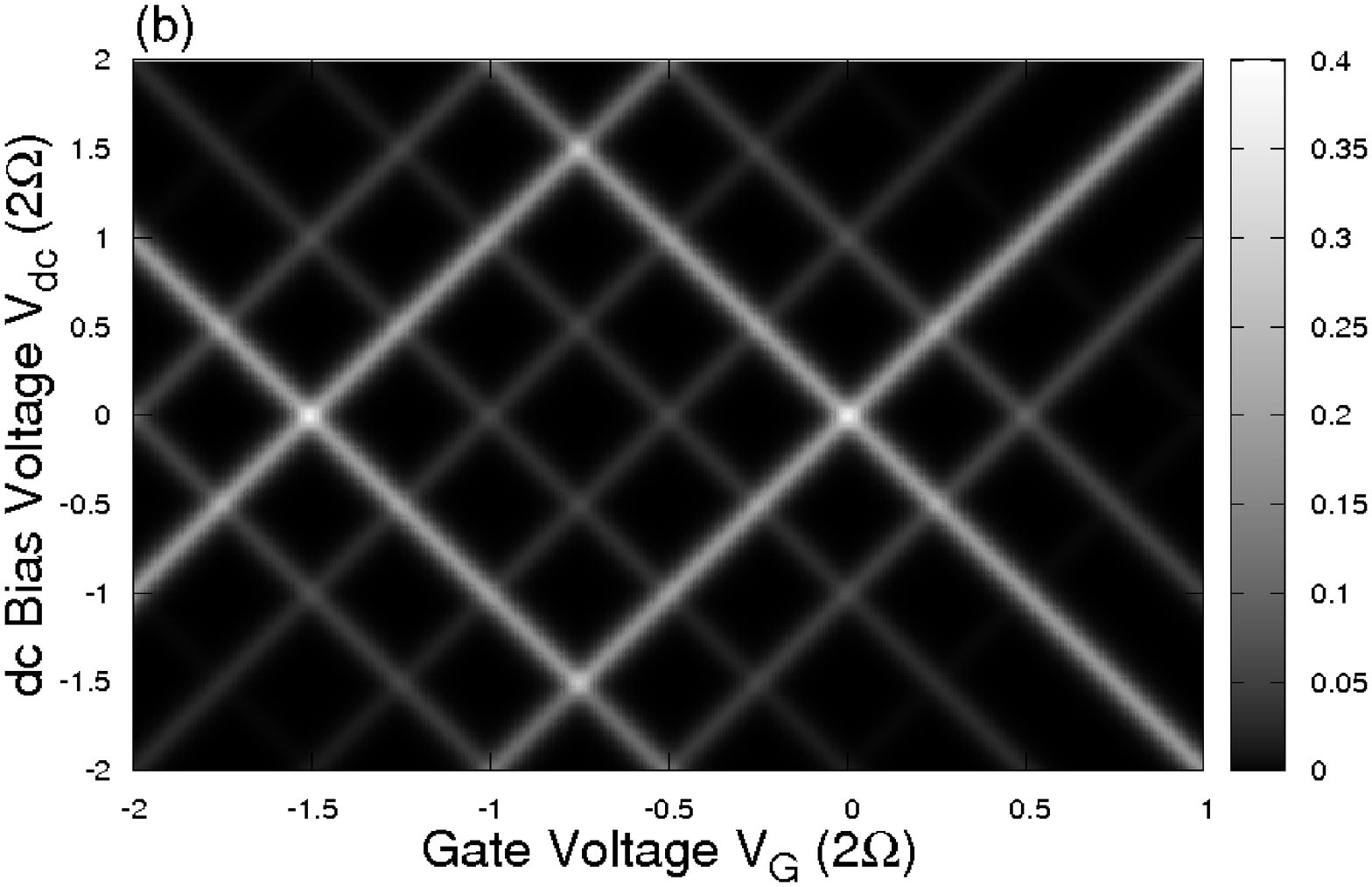} 

\newpage

\includegraphics[angle=0,width=0.8\textwidth]{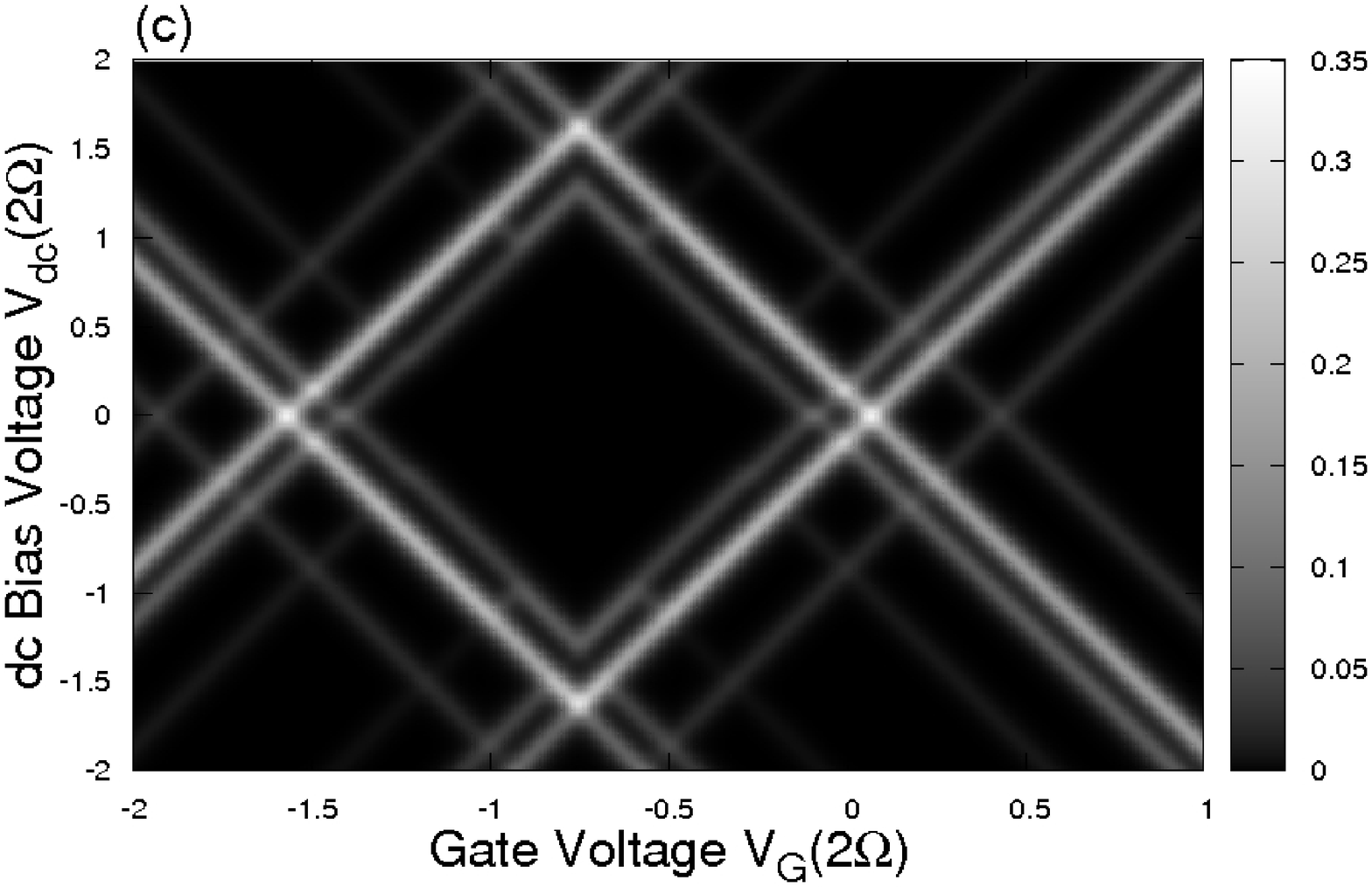} 

\newpage

\end{document}